\documentclass{aa}

\usepackage{natbib}
\usepackage{graphicx}
\usepackage{txfonts}
\usepackage{hyperref}
\usepackage{booktabs}

\newcommand{\numax}{\mbox{$\nu_{\rm max}$}\xspace}
\newcommand{\deltanu}{\mbox{$\langle \Delta\nu \rangle$}\xspace}
\newcommand{\teff}{\mbox{$T_{\rm eff}$}\xspace}
\newcommand{\logg}{\mbox{$\log g$}\xspace}
\newcommand{\feh}{\mbox{$\rm{[Fe/H]}$}\xspace}
\newcommand{\mh}{\mbox{$\rm{[M/H]}$}\xspace}
\newcommand{\afe}{\mbox{$\rm{[\alpha/Fe]}$}\xspace}

\newcommand{\muas}{\mbox{$\mu \rm as$}\xspace}

\newcommand{\kepler}{\emph{Kepler}\xspace}
\newcommand{\gaia}{\emph{Gaia}\xspace}
\newcommand{\ktwo}{K2\xspace}
\newcommand{\tess}{TESS\xspace}

\defcitealias{Lindegren2021}{L21}
\defcitealias{Khan2019}{K19}
\defcitealias{Mosser2011a}{M11}
\defcitealias{Mosser2009}{MA09}
\defcitealias{Elsworth2020}{E20}
\defcitealias{Yu2018}{Y18}

\usepackage[inline]{enumitem}
\usepackage{gensymb}
\usepackage{placeins}
\begin{document}
  \title{Investigating \gaia EDR3 parallax systematics using asteroseismology of Cool Giant Stars observed by \\ \kepler, \ktwo, and \tess}
  \subtitle{I. Asteroseismic distances to 12,500 red-giant stars}
  \titlerunning{Asteroseismic distances to 12,500 red-giant stars}

  \author{S. Khan\inst{1}
        \and
        A. Miglio\inst{2,3,4}
        \and
        E. Willett\inst{4}
        \and
        B. Mosser\inst{5}
        \and
        Y. P. Elsworth\inst{4}
        \and
        R. I. Anderson\inst{1}
        \and
        L. Girardi\inst{6}
        \and
        K. Belkacem\inst{5}
        \and \\
        A. G. A. Brown\inst{7}
        \and
        T. Cantat-Gaudin\inst{8}
        \and
        L. Casagrande\inst{9,10}
        \and 
        G. Clementini\inst{3}
        \and
        A. Vallenari\inst{6}
                }

  \institute{Institute of Physics, Laboratory of Astrophysics, \'Ecole Polytechnique F\'ed\'erale de Lausanne (EPFL), Observatoire de Sauverny, 1290 Versoix, Switzerland\\
              \email{saniya.khan@epfl.ch}
         \and
             Dipartimento di Fisica e Astronomia, Università degli Studi di Bologna, Via Gobetti 93/2, I-40129 Bologna, Italy
         \and
             INAF - Osservatorio di Astrofisica e Scienza dello Spazio di Bologna, Via Gobetti 93/3, I-40129 Bologna, Italy
         \and
         	School of Physics and Astronomy, University of Birmingham,
         	Edgbaston, Birmingham, B15 2TT, UK
         \and
         	LESIA, Observatoire de Paris, PSL Research University, CNRS, Sorbonne Universit\'e, Universit\'e Paris Cit\'e, 92195 Meudon, France
         \and
         	 INAF - Osservatorio Astronomico di Padova, Vicolo dell'Osservatorio 5, I-35122 Padova, Italy
         \and
             Leiden Observatory, Leiden University, Niels Bohrweg 2, 2333 CA Leiden, The Netherlands
         \and
         	 Max-Planck-Institut f\"ur Astronomie, K\"onigstuhl 17, 69117 Heidelberg, Germany
         \and
         	 Research School of Astronomy and Astrophysics, The Australian National University, Canberra, ACT 2611, Australia
         \and
         	 ARC Centre of Excellence for All Sky Astrophysics in 3 Dimensions (ASTRO 3D), Australia
            }

  \date{Received 20 February 2023; accepted 10 April 2023}

  \abstract
  	{\gaia EDR3 has provided unprecedented data that generate a lot of interest in the astrophysical community, despite the fact that systematics affect the reported parallaxes at the level of $\sim 10 \, \rm \mu as$. Independent distance measurements are available from asteroseismology of red-giant stars with measurable parallaxes, whose magnitude and colour ranges more closely reflect those of other stars of interest. In this paper, we determine distances to nearly 12,500 red-giant branch and red clump stars observed by \kepler, \ktwo, and \tess. This is done via a grid-based modelling method, where global asteroseismic observables, constraints on the photospheric chemical composition, and on the unreddened photometry are used as observational inputs. This large catalogue of asteroseismic distances allows us to provide a first comparison with \gaia EDR3 parallaxes. Offset values estimated with asteroseismology show no clear trend with ecliptic latitude or magnitude, and the trend whereby they increase (in absolute terms) as we move towards redder colours is dominated by the brightest stars. The correction model proposed by \citet{Lindegren2021} is not suitable for all the fields considered in this study. We find a good agreement between asteroseismic results and model predictions of the red clump magnitude. We discuss possible trends with the \gaia scan law statistics, and show that two magnitude regimes exist where either asteroseismology or \gaia provides the best precision in parallax.}

  \keywords{asteroseismology --- astrometry --- distance scale --- parallaxes --- stars: distances --- stars: low-mass --- stars: oscillations}

\maketitle


\section{Introduction}
\label{sec:intro}

In December 2020, the early third intermediate data release of \gaia (\gaia EDR3; \citealt{GaiaCollaboration2021}) was published, with updated source list, astrometry, and broad-band photometry in the $G$, $G_{\rm BP}$, and $G_{\rm RP}$ bands. This release represents a significant improvement in both the precision and accuracy of the astrometry and photometry, with respect to \gaia DR2. While quasars yielded a median parallax of $-29 \, \rm \muas$ in DR2, this is now reduced to about $-17 \, \rm \muas$ in \gaia EDR3, with variations at a level of $\sim 10 \, \rm \muas$ depending on position, magnitude and colour \citep{Lindegren2021a}.

With the EDR3 release, \citet{Lindegren2021} (hereafter \citetalias{Lindegren2021}) proposed two offset functions $Z_{\rm 5}(G,\nu_{\rm eff},\beta)$ and $Z_{\rm 6}(G,\hat{\nu}_{\rm eff},\beta)$ applicable to 5- and 6-parameter astrometric solutions, respectively, that give an estimate of the systematics in the parallax measurement as a function of the $G$-band magnitude, effective wavenumber $\nu_{\rm eff}$ or pseudo-colour $\hat{\nu}_{\rm eff}$, and ecliptic latitude $\beta$. Their zero-point correction model is based on quasars, and complemented with indirect methods involving physical binaries and stars in the Large Magellanic Cloud\footnote{Python implementations of both functions are available in the \gaia web pages: \url{https://www.cosmos.esa.int/web/gaia/edr3-code}.}. 

\begin{figure*}
	\centering
	\includegraphics[width=0.75\hsize]{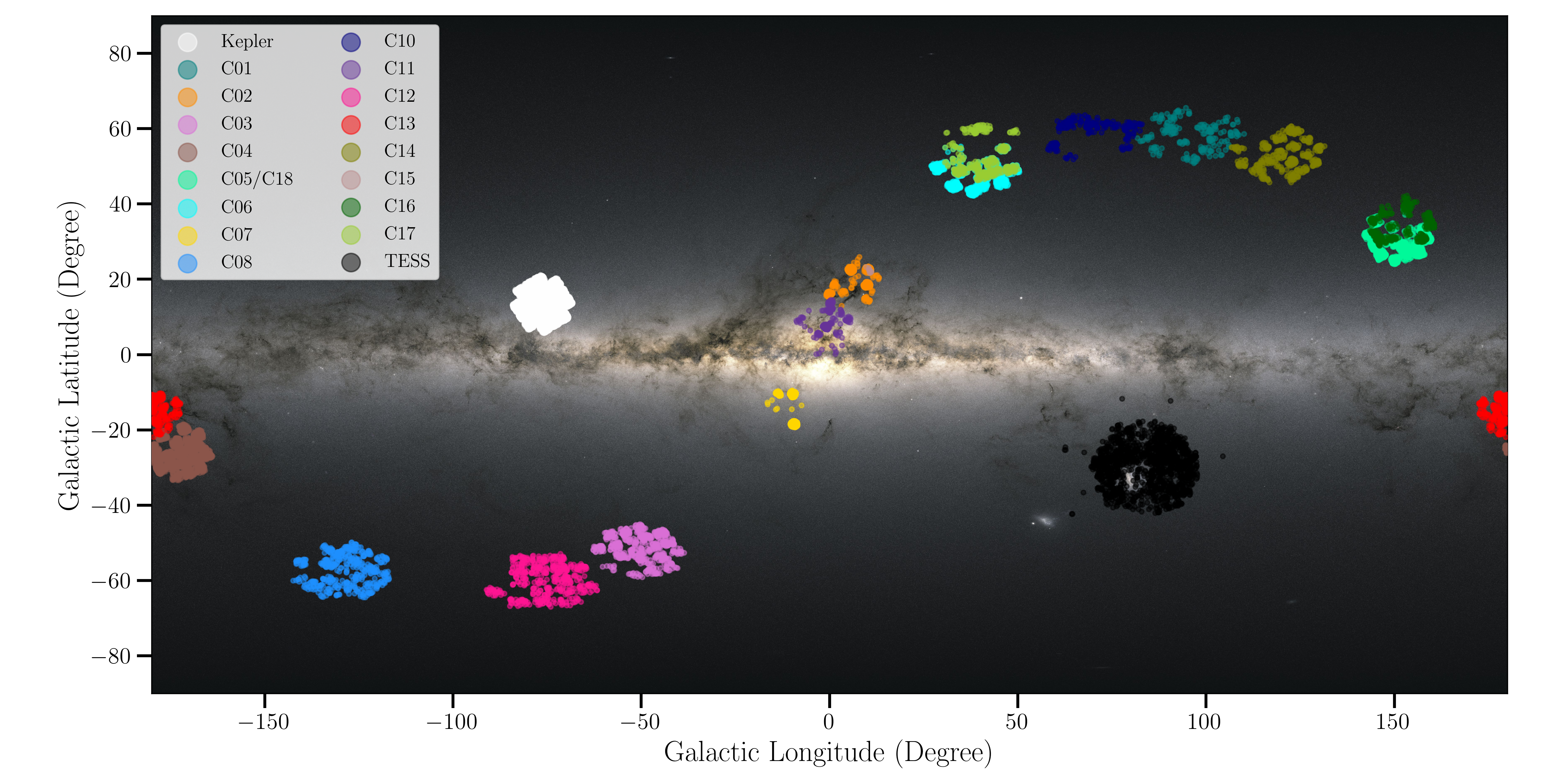}
	\caption{Skymap in Galactic coordinates, showing the location and coverage resulting from the crossmatch between the various asteroseismic fields considered in this study and APOGEE DR17. This figure has been generated using the \texttt{python} package \texttt{mw-plot} (\url{milkyway-plot.readthedocs.io}). The background image comes from ESA/Gaia/DPAC.}
	\label{fig:skymap}
\end{figure*}

Despite the availability of such a correction, \citetalias{Lindegren2021} still encourage users of \gaia EDR3 data to derive their own zero-point estimates, whenever possible. Indeed, some studies dedicated to the comparison between EDR3 parallaxes and independent measurements have found that the inclusion of the \citetalias{Lindegren2021} offset could lead to an over-correction of the parallaxes. All the values reported below give the difference between the corrected EDR3 parallaxes and the other measurements, hence positive values correspond to an over-correction, as a result of applying the \citetalias{Lindegren2021} values\footnote{In this work, we define the residual parallax offset as $\Delta \varpi_{\rm corr} = (\varpi_{\rm EDR3}-Z_5) - \varpi_{\rm other}$, while some of the studies mentioned define it with the opposite sign, i.e. $\Delta \varpi_{\rm corr} = \varpi_{\rm other} - (\varpi_{\rm EDR3}-Z_5)$.} This includes samples based on classical Cepheids ($+14 \pm 5 \, \rm \muas$, \citealt{Riess2021}; $+18 \pm 5 \, \rm \muas$ and $+22 \pm 3 \, \rm \muas$ based on NIR HST and optical \gaia bands respectively, \citealt{CruzReyes2023}; $+22 \pm 4 \, \rm \muas$, \citealt{Molinaro2023}), and RR Lyrae stars \citep[$+22 \pm 2 \, \rm \muas$;][]{Bhardwaj2021}. 
Still, there are other studies which did not report such an overestimation of the parallax zero-point, as can be seen from eclipsing binaries \citep[$-15 \pm 18 \, \rm \muas$;][]{Stassun2021}, red clump stars \citep[$+4.04 \, \rm \muas$, the uncertainty is not reported;][]{Huang2021}, and WUMa-type eclipsing binary systems \citep[$+4.2 \pm 0.5 \, \rm \muas$;][]{Ren2021}.

Following on from our \gaia DR2 study \citep[][hereafter \citetalias{Khan2019}]{Khan2019}, we extend our catalogue of distances using asteroseismic data in the \kepler, \ktwo, and \tess southern continuous viewing zone (\tess-SCVZ) fields, allowing for a first comparison with \gaia EDR3. The asteroseismic and spectroscopic surveys used are briefly described in Sec. \ref{sec:data}. The method for estimating asteroseismology-based parallaxes is introduced in Sec. \ref{sec:seismicplx}. Section \ref{sec:results} presents our parallax zero-point results for \kepler, \ktwo, and \tess separately, and provides a first discussion of global trends seen in ecliptic latitude, magnitude, and effective wavenumber. In Sec. \ref{sec:discussion}, we discuss the magnitude of the red clump as an independent validation of the method, the impact of \gaia scanning law statistics for \ktwo, and the existence of two regimes in magnitude where either the precision of asteroseismology or \gaia dominates. Conclusions are reported in Sec. \ref{sec:conclusions}.

 \begin{table}
	\caption{Overview of the properties of the different datasets: the observation length, range in $G$ magnitude and in $\nu_{\rm eff}$ are given.}
	\label{table:fields}
	\centering
	\begin{tabular}{c | c c c}
		\hline\hline
		Fields & Baseline & $G$ & $\nu_{\rm eff}$ ($\rm \mu m^{-1}$) \\
		\hline
		\kepler & 4 years & [9, 13] & [1.4, 1.5] \\
		\hline
		\ktwo & 80 days & [9, 15] & [1.35, 1.5] \\
		\hline
		\tess-SCVZ & 1 year & [9, 11] & [1.4, 1.5] \\
		\hline
	\end{tabular}
\end{table}

\section{Observational framework}
\label{sec:data}

Our sample is divided into three main parts, summarised in Table \ref{table:fields}, and the location of the various fields is illustrated on Fig. \ref{fig:skymap}. The full datasets with asteroseismic, spectroscopic, and astrometric information are provided along with the paper, and details about the columns are given in App. \ref{app:catalogue}.

\subsection{Asteroseismic information}

We first have first-ascent red-giant branch (RGB) stars and red clump (RC) stars observed by \kepler \citep{Borucki2010}, for which the observation length is the longest: 4 years. The second part of our sample consists of red giants observed by \ktwo, namely \kepler's follow-up mission \citep{Howell2014}. Compared to the two campaigns analysed in \citetalias{Khan2019}, we now have data available for 17 campaigns: C01-08, C10-18. The observations of \ktwo campaigns have a much shorter duration of 80 days. We further analysed very bright ($G < 11$) red-giant stars in the \tess southern continuous viewing zone \citep{Ricker2015}. The \tess full-frame images, from which the asteroseismic data are extracted, are based on 1 year of observations.

For all three surveys, we use the frequency of maximum oscillation power \numax and the average large frequency spacing \deltanu, and consider two different asteroseismic pipelines: \citet[][hereafter \citetalias{Mosser2009}]{Mosser2009} and \citet[][hereafter \citetalias{Elsworth2020}]{Elsworth2020}. We keep stars for which both pipelines return a \numax value in the range [15, 200] $\rm \mu Hz$. Beyond these limits, the \numax estimates are more uncertain and can deviate significantly between \citetalias{Mosser2009} and \citetalias{Elsworth2020}.

\subsection{Spectroscopic information}

For \ktwo, two different surveys are considered for constraints on the photospheric chemical composition, i.e. \teff and \feh (as well as \afe, if available): APOGEE DR17 with near-infrared (NIR) all-sky spectroscopic observations and a resolution of $R \sim 22 500$ \citep{Abdurrouf2022}, and GALAH DR3 with southern hemisphere spectroscopic observations in the optical/NIR and $R \sim 28 000$ \citep{Buder2021}. For \kepler and \tess, we only use APOGEE constraints. Additional flags are also applied following recommendations specific to each spectroscopic survey\footnote{We used the \texttt{STAR\_WARN} and \texttt{STAR\_BAD} flags to clean the APOGEE sample (\url{https://www.sdss.org/dr17/irspec/parameters/}); and \texttt{flag\_sp == 0}, \texttt{flag\_fe\_h == 0}, \texttt{flag\_alpha\_fe == 0} for the GALAH sample (\url{https://www.galah-survey.org/dr3/flags/}).}.

\begin{figure*}
	\centering
	\includegraphics[width=0.70\hsize]{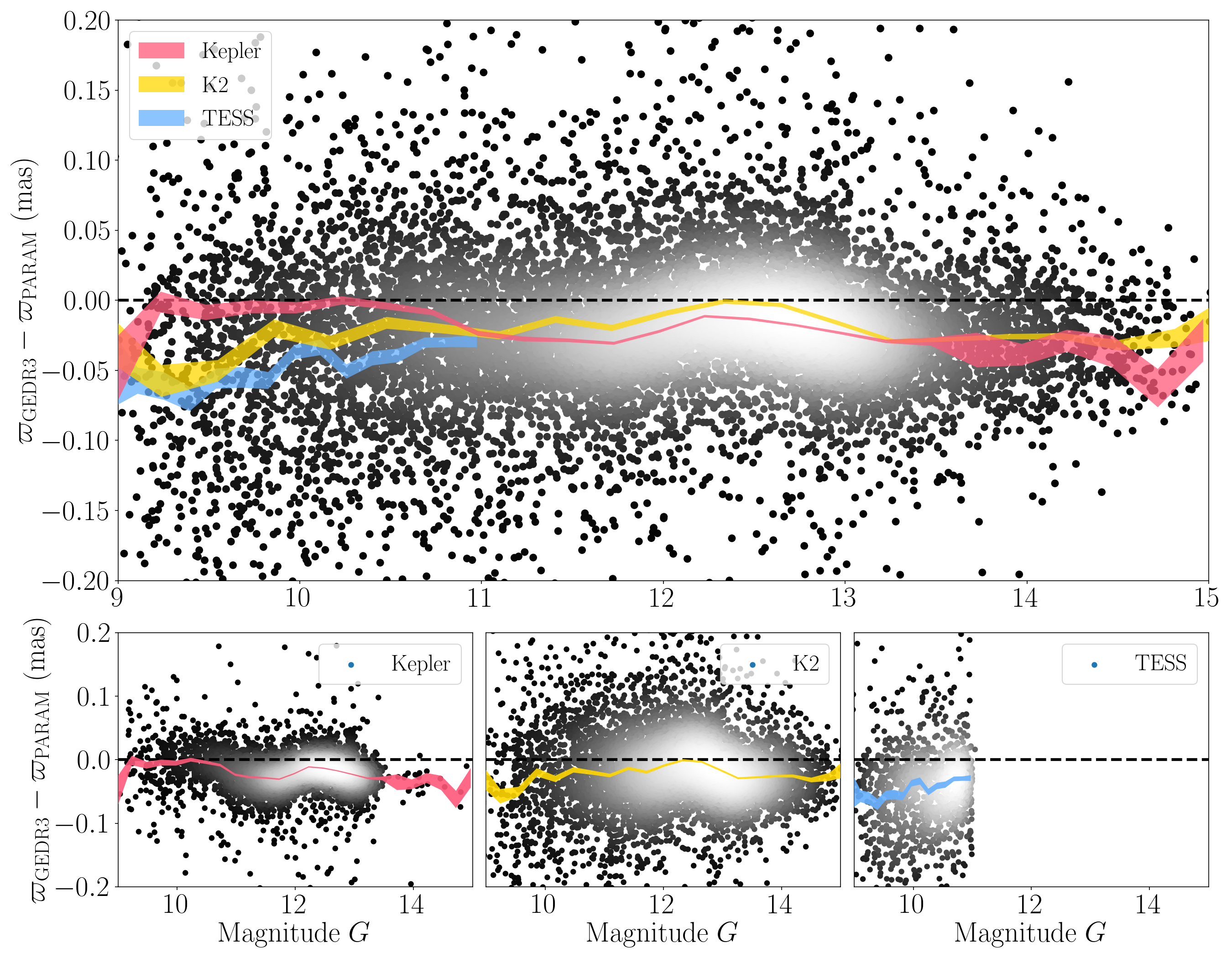}
	\caption{Parallax difference $\varpi_{\rm EDR3}-\varpi_{\rm PARAM}$ as a function of the $G$ magnitude for the full sample (top), \kepler (bottom left), \ktwo (bottom middle), and \tess (bottom right panel), using \citetalias{Elsworth2020} and APOGEE DR17. The colour scale indicates the density of stars, increasing from black to white. The red, yellow, and blue-shaded areas show the median parallax difference binned by magnitude for \kepler, \ktwo, and \tess, respectively.}
	\label{fig:trend_G}
\end{figure*}

\section{Asteroseismic parallaxes}
\label{sec:seismicplx}

Asteroseismic parallaxes are estimated with the Bayesian tool \texttt{PARAM} \citep{Rodrigues2017}. For a given set of observational inputs: \numax, \deltanu, \teff, \logg, \feh, and \afe (when available), as well as photometric measurements, the code will determine the best-fitting stellar parameters by searching among a grid of models. The outputs are given in the form of probability density functions, from which the median and 68\% credible intervals lead to the final parameters of interest and their uncertainties. We refer the reader to \citet{Miglio2021} for a more extensive discussion of the importance of uncertainties related to stellar models.
	
In particular, asteroseismic and spectroscopic constraints are combined together to derive absolute magnitudes in the different passbands, using bolometric corrections from \citet{Girardi2002}. Extinction coefficients are computed adopting the \citet{Cardelli1989} and \citet{ODonnell1994} reddening laws with $R_V = 3.1$. It is then assumed that extinctions in all filters $A_{\lambda}$ are related by a single interstellar extinction curve expressed in terms of its $V$-band value, i.e. $A_{\lambda} (A_V)$. The total extinction $A_V$ and the distance $d$ can then be derived simultaneously. Parallaxes are obtained by inverting the said distances (with a relative uncertainty below $\sim 5\%$), and the error on the distance is propagated to obtain the uncertainty on the parallax. We provide a comparison with \gaia DR3 GSP-Phot distances in App. \ref{app:gsp}.

 \begin{table*}
	\caption{Summary of the different combinations of asteroseismic and spectroscopic constraints used in this study, for the fields considered: \kepler, \ktwo (C01-08, C10-18), and \tess-SCVZ. The median parallax offsets obtained before ($\Delta \varpi = \varpi_{\rm EDR3} - \varpi_{\rm PARAM}$) and after applying \citetalias{Lindegren2021} corrections ($\Delta \varpi_{\rm corr} = \varpi_{\rm EDR3, corr} - \varpi_{\rm PARAM}$) are indicated. For \ktwo, the median offset is calculated considering the 17 campaigns together, and the uncertainty quoted corresponds to the 16th and 84th percentiles. In addition, the full range with the minimum and maximum offsets measured for \ktwo fields is also given.}
	\label{table:data}
	\centering
	\begin{tabular}{c | c c c c c c c}
		\hline\hline
		Fields & Seismo. & Spectro. & $N$ & $\langle \Delta \varpi \rangle$ ($\rm \mu as$) & $\langle \Delta \varpi_{\rm corr} \rangle$ ($\rm \mu as$) & Full range & Full range (corr.) \\
		\hline
		\kepler & \citetalias{Mosser2009} & APOGEE DR17 & 4687 & $-32 \pm 0.4$ & $-12 \pm 0.4$ & - & - \\
		& \citetalias{Elsworth2020} & - & - & $-20 \pm 0.3$ & $-0.4 \pm 0.3$ & - & - \\
		\hline
		\ktwo & \citetalias{Mosser2009} & APOGEE DR17 & 7024 & $-18^{+9}_{-7}$ & $+15^{+6}_{-7}$ & [$-43$, $+2$] & [$-6$, $+33$] \\
		& \citetalias{Elsworth2020} & - & - & $-18^{+12}_{-6}$ & $+14^{+12}_{-7}$ & [$-39$, $+1$] & [$-4$, $+32$] \\
		& \citetalias{Mosser2009} & GALAH DR3 & 5919 & $-19^{+8}_{-6}$ & $+13^{+11}_{-6}$ & [$-68$, $+4$] & [$-41$, $+32$] \\
		& \citetalias{Elsworth2020} & - & - & $-19^{+8}_{-7}$ & $+14^{+12}_{-8}$ & [$-70$, $+15$] & [$-45$, $+43$] \\
		\hline
		\tess-SCVZ & \citetalias{Mosser2009} & APOGEE DR17 & 1253 & $-23 \pm 1.2$ & $-15 \pm 1.2$ & - & - \\
		& \citetalias{Elsworth2020} & - & - & $-41 \pm 1.4$ & $-33 \pm 1.4$ & - & - \\
		\hline
	\end{tabular}
\end{table*}

\section{A first comparison to \gaia EDR3 parallaxes}
\label{sec:results}

To simplify the discussion and figures, we focus on one combination of asteroseismic and spectroscopic constraints. For most \ktwo fields, the offsets measured using \citetalias{Mosser2009} or \citetalias{Elsworth2020}'s seismic observables agree to within a few $\mu$as. For \tess-SCVZ targets, \citet{Mackereth2021} found the \deltanu values returned by \citetalias{Elsworth2020}'s pipeline to be more consistent with individual-mode frequencies (and so to the method employed in models). Hence, we will use the \citetalias{Elsworth2020} asteroseismic pipeline for all three fields. Systematic differences in the spectroscopic parameters published by different surveys affect our results at the level of 5-10 $\rm \mu$as (also partly due to the samples being different). We therefore adopt a single homogeneous spectroscopic dataset with APOGEE DR17 to ensure greatest precision.

A summary of the parallax zero-points derived is given in Table \ref{table:data}, while individual offsets for all combinations of seismic and spectroscopic constraints are provided in App. \ref{app:table}. More detailed checks on how the asteroseismic method and the choice of spectroscopy affect the analysis of \gaia systematics will be presented in a forthcoming paper (Khan et al., in prep.).

\subsection{Separate analyses for \kepler, \ktwo, and \tess}
\label{sec:individual}

In the following, our results are based on 5-parameter astrometric solutions only. We estimate the parallax offset for each field, before and after applying \citetalias{Lindegren2021} corrections to \gaia parallaxes, and study potential trends with asteroseismic, spectroscopic, and photometric parameters. We had initially compared our results with \citet{Zinn2021}'s analysis of \kepler targets. However, by doing so, we noticed a misuse in the \citetalias{Lindegren2021} corrections computed in their study, that is they used $\sin \beta$ instead of $\beta$ in the Python code.\\

We investigate the parallax difference $\Delta \varpi = \varpi_{\rm EDR3}-\varpi_{\rm PARAM}$ as a function of $G$, and verify that $\Delta \varpi$ is negative for all fields, in the sense that \gaia parallaxes are smaller (Fig. \ref{fig:trend_G}). We apply the same analysis as in \citetalias{Khan2019} on the \kepler sample, but this time with \gaia EDR3 parallaxes and updated APOGEE constraints. $\Delta \varpi$ shows fairly flat trends as a function of the ecliptic latitude, the effective wavenumber, the frequency of maximum oscillation, the mass inferred from \texttt{PARAM}, and the metallicity, but not for the $G$ magnitude which displays a non-linear feature (see bottom left panel of Fig. \ref{fig:trend_G}). This relation with $G$ is expected due to changes in the gating scheme or in the window size (see Fig. 17 in \citealt{Fabricius2021}). Despite the larger scatter and a higher proportion of fainter stars compared to \kepler, we also observe a non-linear trend as a function of $G$ if we combine all \ktwo fields together, which have an ecliptic latitude near zero (see bottom middle panel of Fig. \ref{fig:trend_G}). However, our \tess-SCVZ sample is too bright to see this trend.

Figure \ref{fig:trend_G_L21} is similar to Fig. \ref{fig:trend_G}, but shows instead the parallax offset residuals $\Delta \varpi_{\rm corr} = \varpi_{\rm EDR3, corr}-\varpi_{\rm PARAM}$, with $Z_5$-corrected \gaia EDR3 parallaxes. This removes the non-linear trend with $G$. It is also clear from Fig. \ref{fig:trend_G_L21} that \citetalias{Lindegren2021} corrections underestimate the parallax offset in the case of \tess, and overestimate it when it comes to \ktwo fields. But in \kepler, the residual parallax offset gets very close to zero. This suggests that \citetalias{Lindegren2021} corrections are not universally suited for different types of sources, spanning a wide range of positions, magnitudes, and colours.

For some of the \ktwo campaigns (and independently of the spectroscopy used), we notice a significant trend of the parallax difference with the stellar mass. As we do not observe such a trend with mass for \kepler and \tess, we suspect that it could be related to, e.g., different noise levels in the various \ktwo campaigns. We tried using scaling relations to compute the mass and the asteroseismic parallax instead of \texttt{PARAM}, tested different asteroseismic pipelines and spectroscopic surveys, and removed high \numax stars. Unfortunately, none of these made a difference and this is still being investigated (by BM and YE), as it might directly be related to the accuracy of seismically-inferred parameters. 

\subsection{A global picture}
\label{sec:global}

\begin{figure}
	\centering
	\includegraphics[width=\hsize]{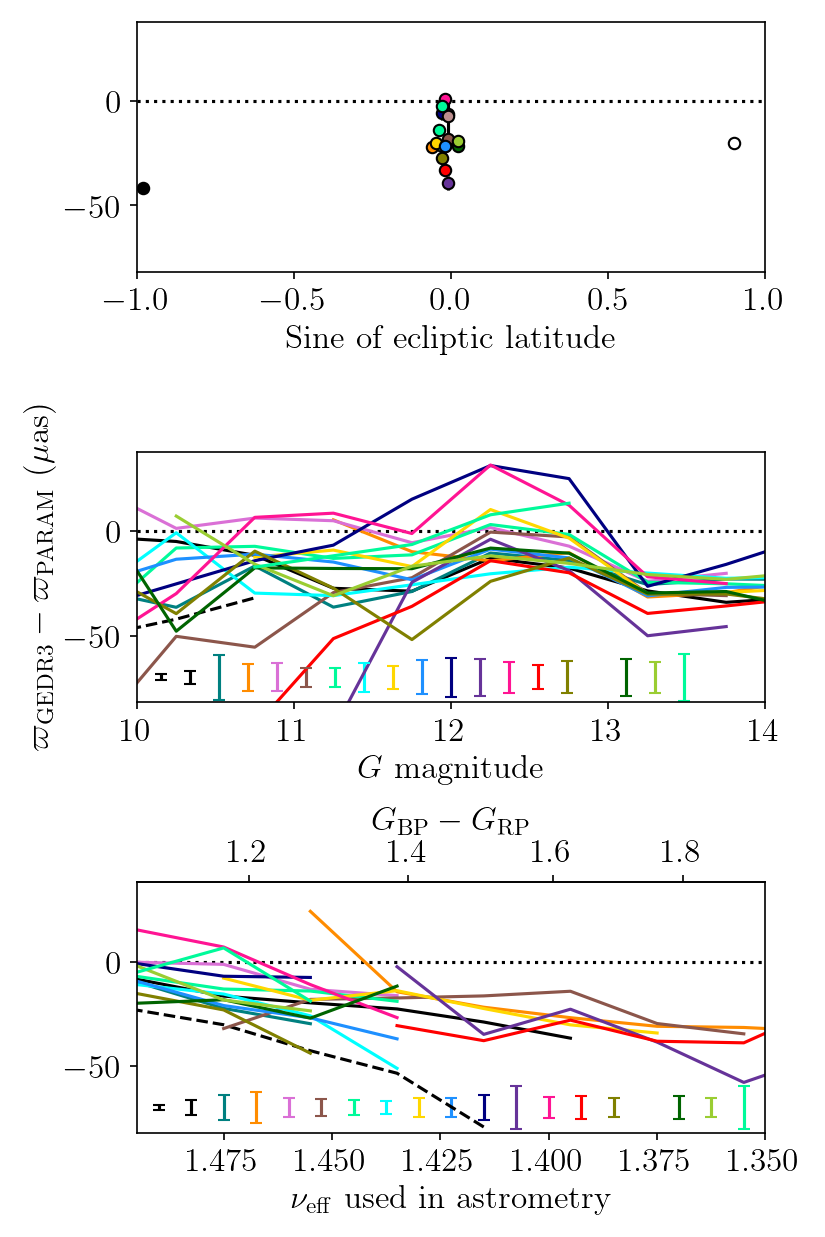}
	\caption{\textit{Top}: Median parallax offsets as estimated from asteroseismology (\citetalias{Elsworth2020}+APOGEE), as a function of the sine of ecliptic latitude. \kepler and \tess are plotted as white and black symbols, respectively. The coloured symbols correspond to the various \ktwo fields, and follow the colour scheme adopted in Fig. \ref{fig:skymap}. \textit{Middle and bottom}: Median parallax difference binned by $G$ magnitude (middle) and effective wavenumber (bottom panel). \kepler and \tess are plotted as black solid and dashed lines, respectively. The median uncertainty on the parallax difference is shown in the lower part of each panel. C15 does not appear in the two bottom panels as there are not enough stars to bin in $G$ and $\nu_{\rm eff}$.}
	\label{fig:offset_summary}
\end{figure}

In Fig. \ref{fig:offset_summary}, we show the offset estimates $\Delta \varpi$ suggested from the difference between the uncorrected \gaia EDR3 and \texttt{PARAM} parallaxes, in the \kepler, the individual \ktwo campaigns, and the \tess-SCVZ fields. We analyse the relation between the parallax zero-point and the ecliptic latitude $\beta$, the $G$ magnitude, and the effective wavenumber $\nu_{\rm eff}$, which are the three parameters constituting the \citetalias{Lindegren2021} correction model. 

We first note that the offsets measured from asteroseismology either are close to zero or negative, and lie (at most) a few tens of $\mu$as away from the zero-point suggested by quasars ($\sim -17 \, \rm \mu as$). All the \ktwo campaigns have similar $\sin \beta$, close to zero, which is expected as the \ktwo survey observed solar-like oscillators all along the ecliptic. 

For individual K2 campaigns, the parallax difference also follows a non-linear relation with $G$, in line with what was discussed in Sec. \ref{sec:individual}. The bottom panel of Fig. \ref{fig:offset_summary} suggests that the parallax difference becomes more negative as we go towards lower $\nu_{\rm eff}$, i.e. redder colours. This is also apparent for $\nu_{\rm eff} \lesssim 1.40$, where we have fewer campaigns. But one has to keep in mind that this trend is dominated by bright stars, for which other caveats exist (see e.g. Sec. \ref{sec:bias}), which tend to drag the parallax difference towards substantially negative values (as can be seen from the middle panel of Fig. \ref{fig:offset_summary}).

\section{Discussion}
\label{sec:discussion}

\subsection{Magnitude of the red clump}

\begin{figure}
	\centering
	\includegraphics[width=\hsize]{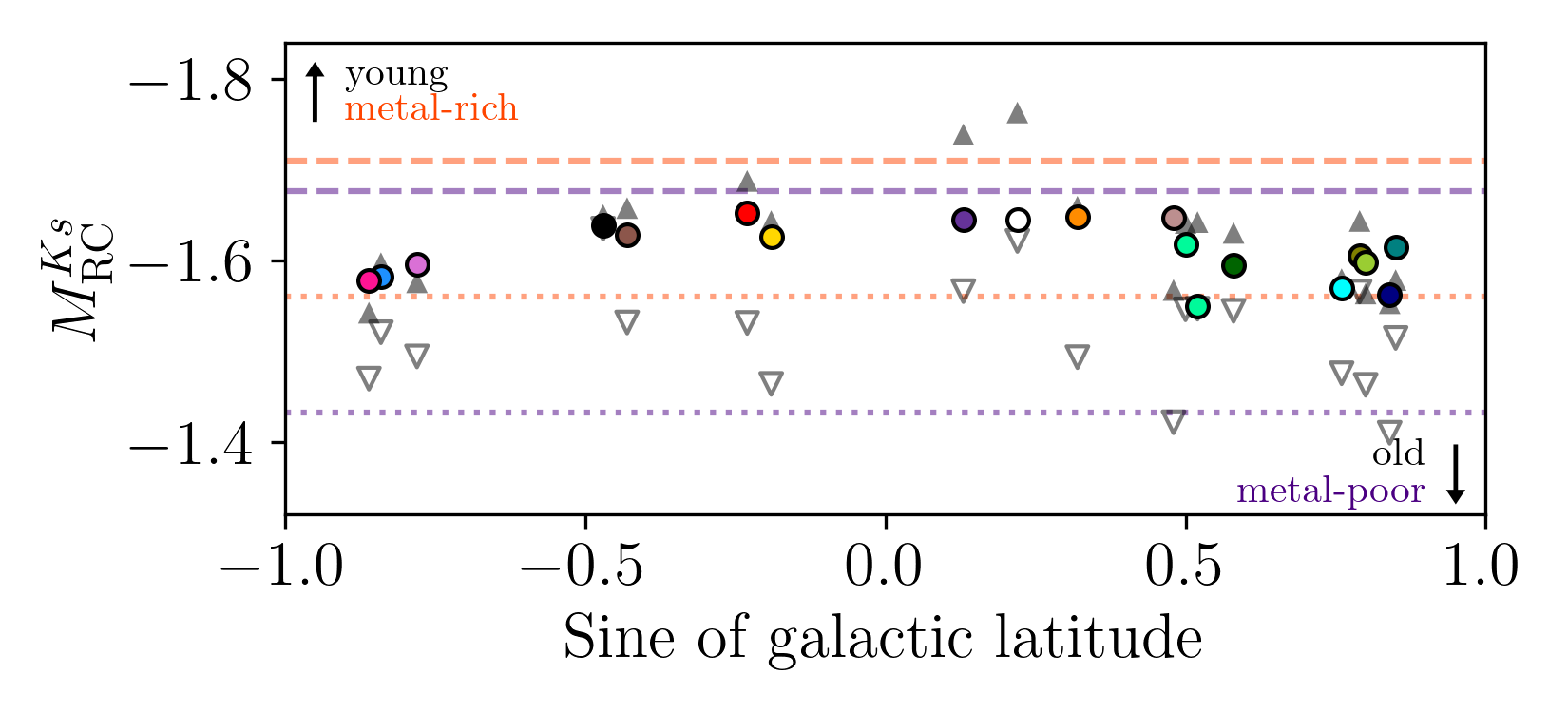}
	\caption{Magnitude of the red clump in the $K_s$ band, as estimated from our asteroseismic sample (circles, same colour scheme as in Fig. \ref{fig:skymap}), \gaia EDR3 samples before (filled triangles) and after applying \citetalias{Lindegren2021} corrections (open triangles), as a function of the sine of the galactic latitude. The lines show predictions from modelling: purple/orange for a metal-poor ($-0.4$ dex)/metal-rich model ($+0.2$ dex); dashed/dotted for a young (5 Gyr)/old model \citep[12 Gyr; see Fig. 8 in][]{Girardi2016}. These values have been chosen to be representative of the lower and upper bounds of the metallicity and age distributions in the asteroseismic fields.}
	\label{fig:rcmag}
\end{figure}

As a way to validate our findings, we also analyse the information provided by the magnitude of the red clump. In Fig. \ref{fig:rcmag}, we show different estimates of the absolute magnitude of the clump as a function of the galactic latitude $b$. The first estimate is based on the $K_s$-band absolute magnitude computed by \texttt{PARAM}, which is thus representative of our asteroseismic samples. For the other two estimates, we select \gaia EDR3 sources centred around the coordinates of each field and with $1 < G < 15$: one estimate is calculated using the inverted \gaia uncorrected parallaxes, and the other with corrected parallaxes (using the \citetalias{Lindegren2021} correction model). In order to be able to safely use inverted parallaxes, we restrict our samples to \gaia sources with a relative parallax uncertainty lower than 10\%. Extinctions are calculated with the combined map \citep{Marshall2006,Green2019,Drimmel2003} from \texttt{mwdust}\footnote{\url{https://github.com/jobovy/mwdust}} \citep{Bovy2016}, and should only have a minor effect on the current analysis as we are working with $K_s$-band magnitudes. For each dataset, we then compute the mode of the magnitude of the red clump using a Kernel Density Estimation with a fixed bandwidth (equal to 0.1) on the corresponding histogram.

The magnitude of the red clump shows a trend with the galactic latitude. Figure 3 of \citet{Ren2021} shows that the parallax offset is observed to be more negative for $\sin b \sim 0$, which could explain why the filled triangles are more luminous in our Fig. \ref{fig:rcmag}. On the other hand, a brighter red clump luminosity would result from a younger and more metal-rich population. This trend is visible when using the seismic sample or the \gaia EDR3 sample without applying \citetalias{Lindegren2021} parallax corrections. But the corrected \gaia EDR3 sample shows a flat trend instead, which again supports the idea that the \citetalias{Lindegren2021} zero-point model is not suited to every kind of star (see also Sec. \ref{sec:individual}). In addition, results from asteroseismology agree well with model predictions \citep[see e.g.][]{Girardi2016}.

\subsection{Impact of \gaia scanning law statistics for \ktwo}

\begin{figure}
	\centering
	\includegraphics[width=0.85\hsize]{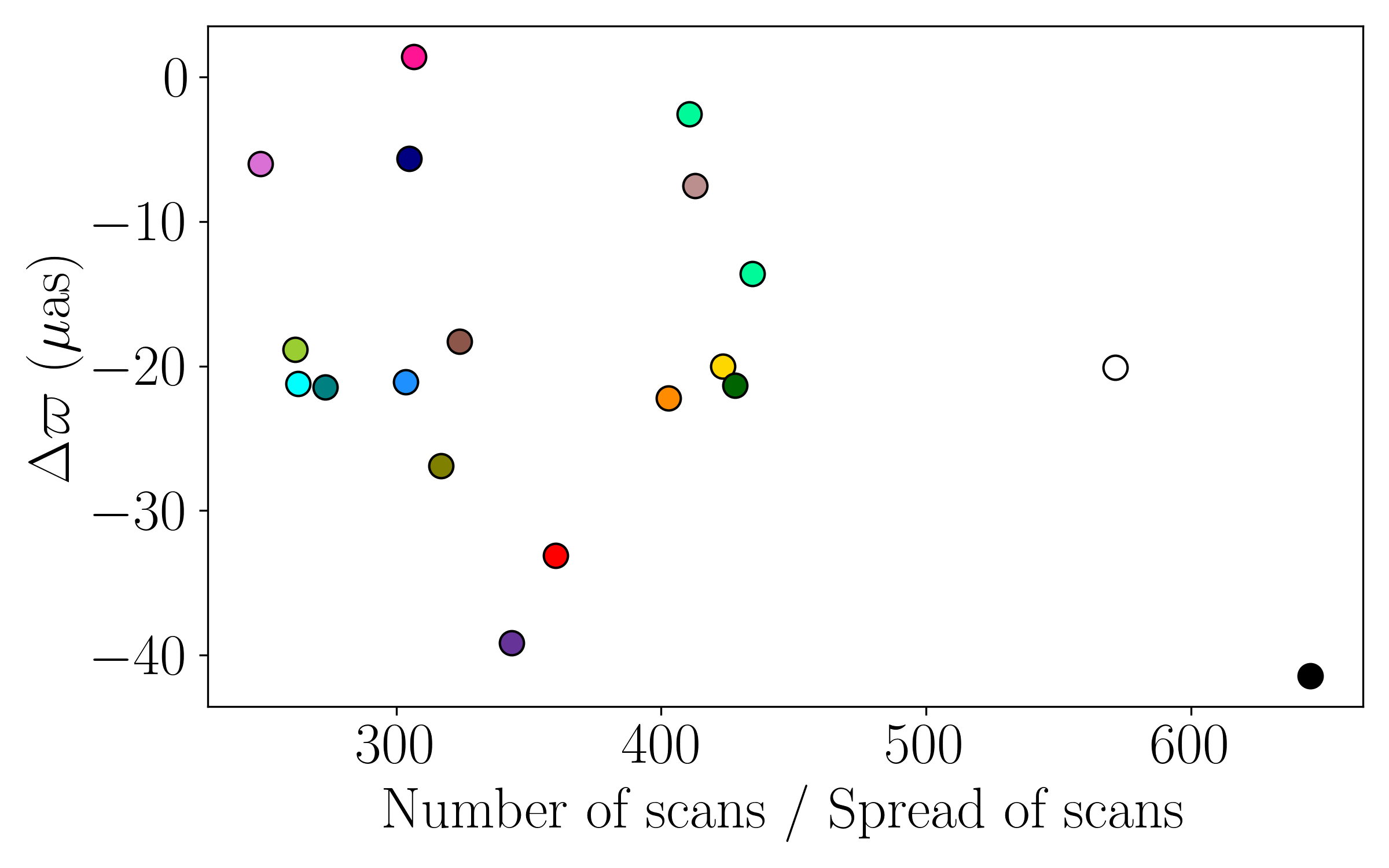}
	\caption{Parallax offset, measured as the difference between uncorrected \gaia EDR3 and asteroseismic parallaxes, as a function of the number of scans over the spread of scans. A low value suggests fewer scans clustered at one time of the year, hence an astrometry of lesser quality, while a higher value corresponds to a greater number of scans better separated in time. The colour scheme is the same as in Fig. \ref{fig:skymap}.}
	\label{fig:scanlaw}
\end{figure}

We look into whether the spread in parallax zero-points suggested by the \ktwo fields could be related to \gaia scan law statistics. For this, we extracted both the average number of scans and spread of scans throughout the year for \gaia EDR3 \citep[see Fig. 1 of][for the all-sky distribution of these quantities in \gaia DR2]{Everall2021}. The high ecliptic latitude fields, \kepler and \tess, show a high number of scans and an important spread of scans. On the other hand, for \ktwo we find fewer scans that are often concentrated at a single time of the year, which is consistent with the fact that these fields are located unfavourably with respect to the \gaia scanning law. As a result, the uncertainty on \gaia EDR3 parallaxes is larger for \ktwo, compared to \kepler and \tess. Apart from these obvious differences, we do not observe any trend of the parallax offset with the scan law statistics, between the various \ktwo campaigns (see Fig. \ref{fig:scanlaw}).

\subsection{Existence of two magnitude regimes}
\label{sec:bias}

\begin{figure}
	\centering
	\includegraphics[width=\hsize]{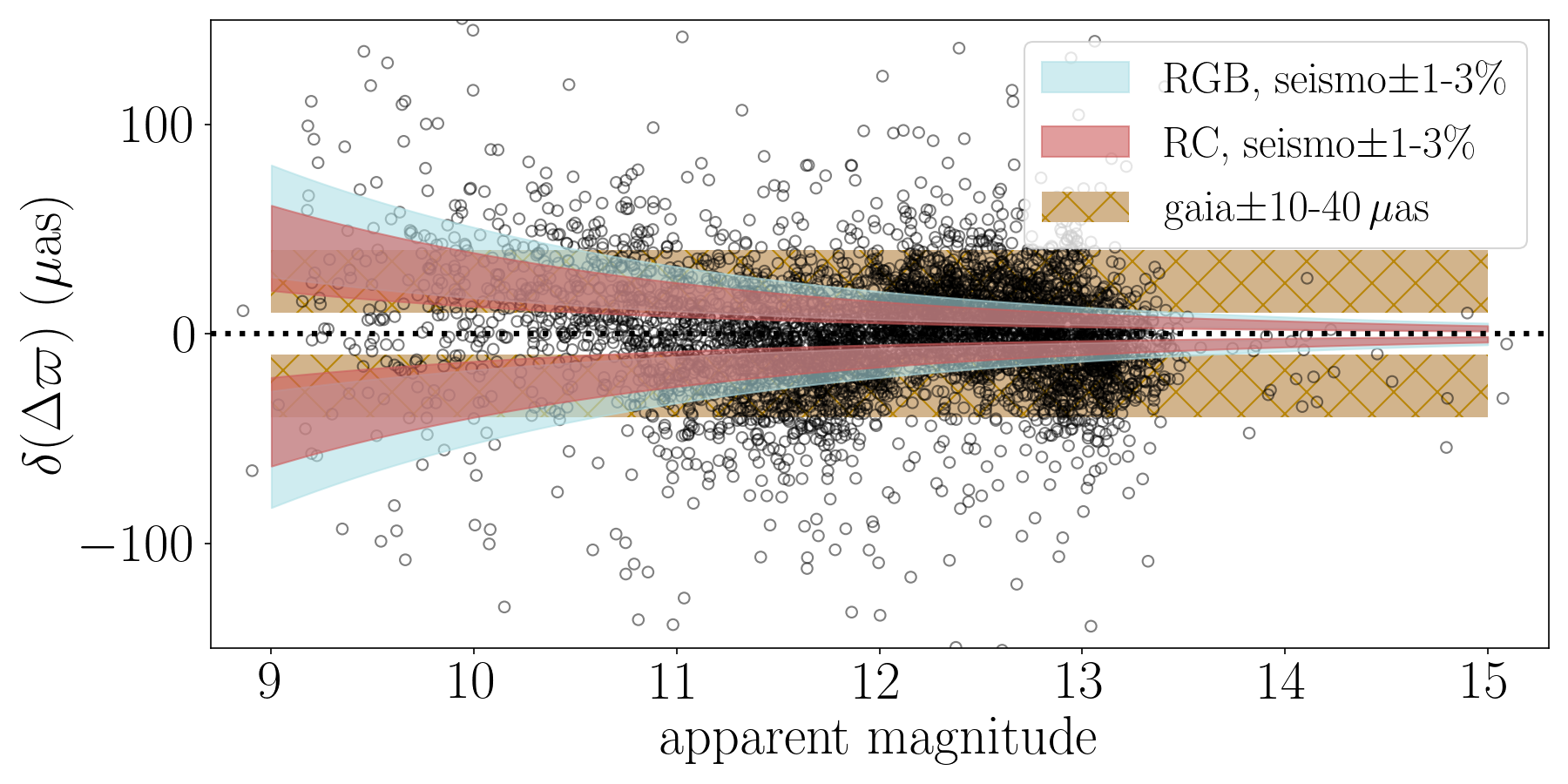}
	\caption{Bias in the parallax difference as a function of the apparent magnitude. Two mock stars are considered for the asteroseismic bias: an RGB star (blue) and a RC star (red; see text for details). We show the asteroseismic bias that would result from a 1-3\% systematic uncertainty in radius. The \gaia bias, $\pm 10$-$40 \, \rm \muas$ in parallax, is shown as a yellow hatched region. \kepler observations are shown in the background, after subtracting the mean parallax offset.}
	\label{fig:systematics}
\end{figure}

Figure \ref{fig:systematics} illustrates the biases arising from asteroseismology or \gaia's side, as a function of the $G$-band apparent magnitude. The asteroseismic bias corresponds to a fractional systematic uncertainty in radius, hence in distance; while the \gaia bias would be related to the effect of a systematic (absolute) uncertainty in parallax.

In order to test this, we consider two mock stars: one RGB star with $L = 30\, \rm L_{\rm \odot}$, $T_{\rm eff} = 4630 \, \rm K$, [Fe/H] = 0.0 dex, $\log g = 2.6$, and a RC star with $L = 50\, \rm L_{\rm \odot}$, $T_{\rm eff} = 4740 \, \rm K$, [Fe/H] = 0.0 dex, $\log g = 2.4$. We then estimate the absolute magnitude in $G$-band. We consider a range of apparent magnitude values [9, 15], and compute a parallax value for each magnitude. "Biased" parallaxes are then estimated: either adding a constant to the distance modulus, which would correspond to a fractional uncertainty in radius (asteroseismic bias); or adding a constant to the parallax itself (\gaia bias). For the former, we consider a $\pm$1-3\% bias in radius, which corresponds to the 16th and 84th percentiles for the \kepler dataset; while for the latter, we use a range of $\pm 10$-$40 \, \muas$.

We show on Fig. \ref{fig:systematics} how such biases may affect the estimation of the parallax zero-point. The \kepler dataset is also shown in the background (after subtracting the mean parallax offset), to see how the order of magnitude of these biases compare with the actual observations. The existence of two regimes becomes quite clear: at the bright end, the comparisons in terms of parallax difference are dominated by systematics affecting the seismic parallax; and at the faint end, systematics from asteroseismology are much less dominant and one can potentially expose \gaia's. 
This division stems from the fact that the fractional uncertainty on asteroseismic distances (or parallaxes) is largely distance independent, but the absolute precision (in pc or mas) is very much distance dependent, so it becomes worse than \gaia's in nearby objects.

\section{Conclusions}
\label{sec:conclusions}

We carried out a follow-up of our 2019 study \citep{Khan2019} to investigate the \gaia EDR3 parallax zero-point, for a significantly larger number of asteroseismic fields: \kepler, 17 \ktwo campaigns, and the \tess-SCVZ. Our analysis is similar to that of \citet{Zinn2021} for the \kepler field but goes beyond with the addition of \ktwo and \tess, also making sure that we combine asteroseismic and spectroscopic constraints in a fully homogeneous way. This has the benefit of exploring \gaia parallax systematics for the same type of objects but with a wide range of positions over the sky within a single study. A quick comparison of asteroseismic distances with \gaia DR3's GSP-Phot estimates shows that a reasonable agreement is found for objects within 2 kpc.

First, we confirm the positional dependence of the \gaia parallax zero-point: \kepler has an offset of $\sim -20 \, \mu$as, \ktwo campaigns span a wide range between $\sim -39$ and $+1 \, \mu$as, and \tess shows an offset of $\sim -41 \, \mu$as when using \citetalias{Elsworth2020} and APOGEE constraints. 

The inclusion of the \citet{Lindegren2021} zero-point estimates improves the agreement between \gaia and asteroseismology in the case of \kepler and, to a much lesser extent, \tess. However, in most \ktwo fields, it can significantly over-correct the parallax difference, sometimes resulting in large positive parallax offsets. This underlines the need to consistently determine the parallax systematics applicable to the sample of interest, taking into account the distributions in position, magnitude, and colour. Such an over-correction had already been suggested by former studies \citep[e.g.][]{Bhardwaj2021,Riess2021}.

Lastly, in terms of magnitude and colour dependence, we show that asteroseismology provides us with strong constraints on the \gaia EDR3 parallax zero-point, in ranges that are not necessarily well-sampled by \citetalias{Lindegren2021} corrections. There are no clear trends with the ecliptic latitude or the $G$ magnitude, but the zero-point values tend to increase (in absolute terms) towards redder colours (lower $\nu_{\rm eff}$). Although this trend seems to be dominated by caveats associated with stars at brighter magnitudes. Moreover, we find that seismic-based estimates of the red clump magnitude are consistent with theoretical predictions of $M^{Ks}_{\rm RC}$, and that the inclusion of the \citetalias{Lindegren2021} offset tends to make the red clump too faint. We do not find any correlation between \gaia scan law statistics and parallax offset estimates for the \ktwo fields. We also use two mock stars to illustrate the existence of two regimes: bright magnitudes, where \gaia's precision is better than asteroseismology's; and faint magnitudes, where we can expose \gaia's limits thanks to seismology's precision.

With this study, we present asteroseismology as a powerful tool for constraining \gaia systematics. Red giants come with several benefits: they are single stars with measurable parallaxes and without large-amplitude photometric variations, and consequently differ substantially from eclipsing binaries, quasars, RR Lyrae and Cepheids. Further progress is foreseen with \gaia DR4 which will have improved parallax uncertainties and reduced systematics. And, in a forthcoming paper, we will look in more detail at the uncertainties potentially affecting parallax estimates from asteroseismology and \gaia, and see how we can define the best sample to investigate parallax systematics in \gaia.

\begin{acknowledgements}
This work has made use of data from the European Space Agency (ESA) mission
{\it Gaia} (\url{https://www.cosmos.esa.int/gaia}), processed by the {\it Gaia} Data Processing and Analysis Consortium (DPAC, \url{https://www.cosmos.esa.int/web/gaia/dpac/consortium}). Funding for the DPAC has been provided by national institutions, in particular the institutions participating in the {\it Gaia} Multilateral Agreement. RIA and SK are funded by the Swiss National Science Foundation (SNSF) through an Eccellenza Professorial Fellowship
(award PCEFP2\_194638). AM and EW acknowledge support from the ERC Consolidator Grant funding scheme ({\em project ASTEROCHRONOMETRY}, G.A. n. 772293). This research was supported by the International Space Science Institute (ISSI) in Bern, through ISSI International Team project \#490,  SHoT: The Stellar Path to the Ho Tension in the Gaia, TESS, LSST and JWST Era.
\end{acknowledgements}

\bibliographystyle{aa} 
\bibliography{references} 

\begin{appendix}

\section{Catalogues of asteroseismic, spectroscopic, and astrometric properties for \kepler, \ktwo, and \tess red giants}
\label{app:catalogue}

\begin{table*}
	\caption{Description of the columns contained in the catalogues we release in this work. There are four datasets in total: \kepler+ APOGEE, \ktwo+ APOGEE, \ktwo+ GALAH, and \tess+ APOGEE. In these tables, we compile asteroseismic information from \citetalias{Mosser2009} and \citetalias{Elsworth2020}, spectroscopic constraints from APOGEE DR17 \citep{Abdurrouf2022} or GALAH DR3 \citep{Buder2021}, and astrometric properties from \gaia (E)DR3 \citep{GaiaCollaboration2021,GaiaCollaboration2022}. Where relevant, uncertainties defined either as the standard deviation (`\texttt{err}') or as the 16/84th percentiles (`\texttt{68L}', `\texttt{68U}') are provided. For APOGEE, we adopt conservative uncertainty values of 50 K and 0.05 dex for \teff and \mh, but keep the original values if they are larger. For GALAH, if \afe is not available, then the value and uncertainty on \mh are simply equal to those of \feh. \label{table:catalogue}}
	\begin{tabular}{llc}
		\hline
		Column Identifier & Description & Units \\
		\hline
		\texttt{KIC}/\texttt{K2\_ID}/\texttt{TIC} & \kepler/\ktwo/\tess ID & $\mathrm{None}$ \\
		\texttt{K2\_campaign} & \ktwo campaign number & $\mathrm{None}$ \\
		\texttt{GEDR3\_source\_id} & \gaia EDR3 source id & $\mathrm{None}$ \\
		\texttt{GEDR3\_ra} & \gaia EDR3 right ascension & $\mathrm{deg}$ \\
		\texttt{GEDR3\_dec} & \gaia EDR3 declination & $\mathrm{deg}$ \\
		\texttt{GEDR3\_l} & \gaia EDR3 galactic longitude & $\mathrm{deg}$ \\
		\texttt{GEDR3\_b} & \gaia EDR3 galactic latitude & $\mathrm{deg}$ \\
		\texttt{GEDR3\_ecl\_lon} & \gaia EDR3 ecliptic longitude & $\mathrm{deg}$ \\
		\texttt{GEDR3\_ecl\_lat} & \gaia EDR3 ecliptic latitude & $\mathrm{deg}$ \\
		\texttt{GEDR3\_parallax} & \gaia EDR3 parallax & $\mathrm{mas}$ \\
		\texttt{GEDR3\_parallax\_Z5} & \citetalias{Lindegren2021} correction (5p) & $\mathrm{mas}$ \\
		\texttt{GEDR3\_phot\_g\_mean\_mag} & \gaia EDR3 $G$-band magnitude & $\mathrm{mag}$ \\
		\texttt{GEDR3\_phot\_bp\_mean\_mag} & \gaia EDR3 $G_{BP}$-band magnitude & $\mathrm{mag}$ \\
		\texttt{GEDR3\_phot\_rp\_mean\_mag} & \gaia EDR3 $G_{RP}$-band magnitude & $\mathrm{mag}$ \\
		\texttt{GEDR3\_nu\_eff\_used\_in\_astrometry} & \gaia EDR3 effective wavenumber & $\mathrm{\mu m^{-1}}$ \\
		\texttt{GEDR3\_pseudocolour} & \gaia EDR3 pseudocolour & $\mathrm{\mu m^{-1}}$ \\
		\texttt{GEDR3\_astrometric\_params\_solved} & \gaia EDR3 number of astrometric parameters solved & $\mathrm{None}$ \\
		\texttt{GEDR3\_ruwe} & \gaia EDR3 renormalised unit weight error & $\mathrm{None}$ \\
		\texttt{GEDR3\_ipd\_frac\_multi\_peak} & \gaia EDR3 percent of successful-IPD windows with more than one peak & $\mathrm{None}$ \\
		\texttt{GEDR3\_ipd\_gof\_harmonic\_amplitude} & \gaia EDR3 amplitude of the IPD GoF versus position angle of scan & $\mathrm{None}$ \\
		\texttt{GDR3\_non\_single\_star} & \gaia DR3 flag indicating the availability in Non-Single Star tables & $\mathrm{None}$ \\
		\texttt{GDR3\_distance\_gspphot} & \gaia DR3 distance from GSP-Phot & $\mathrm{pc}$ \\
		\texttt{GDR3\_ag\_gspphot} & \gaia DR3 extinction in $G$-band from GSP-Phot & $\mathrm{mag}$ \\
		\texttt{2MASS\_kmag} & 2MASS $K_{S}$-band magnitude & $\mathrm{mag}$ \\
		\texttt{APOGEE\_ID} & APOGEE ID & $\mathrm{None}$ \\
		\texttt{APOGEE\_teff} & APOGEE $T_{\mathrm{eff}}$ & $\mathrm{K}$ \\
		\texttt{APOGEE\_feh} & APOGEE $\mathrm{[Fe/H]}$ & $\mathrm{None}$ \\
		\texttt{APOGEE\_alpha} & APOGEE $\mathrm{[\alpha/Fe]}$ & $\mathrm{None}$ \\
		\texttt{APOGEE\_mh} & APOGEE $\mathrm{[M/H]}$ & $\mathrm{None}$ \\
		\texttt{APOGEE\_ASPCAPFLAGS} & APOGEE flag for issues associated with the ASPCAP fits & $\mathrm{None}$ \\
		\texttt{GALAH\_teff} & GALAH $T_{\mathrm{eff}}$ & $\mathrm{K}$ \\
		\texttt{GALAH\_feh} & GALAH $\mathrm{[Fe/H]}$ & $\mathrm{None}$ \\
		\texttt{GALAH\_alpha} & GALAH $\mathrm{[\alpha/Fe]}$ & $\mathrm{None}$ \\
		\texttt{GALAH\_mh} & GALAH $\mathrm{[M/H]}$ & $\mathrm{None}$ \\
		\texttt{GALAH\_flag\_sp} & GALAH stellar parameter quality flag & $\mathrm{None}$ \\
		\texttt{GALAH\_flag\_fe\_h} & GALAH $\mathrm{[Fe/H]}$ quality flag & $\mathrm{None}$ \\
		\texttt{GALAH\_flag\_alpha\_fe} & GALAH $\mathrm{[\alpha/Fe]}$ quality flag & $\mathrm{None}$ \\
		\texttt{MA09\_numax} & $\nu_{\mathrm{max}}$ from \citetalias{Mosser2009} pipeline & $\mathrm{\mu Hz}$ \\
		\texttt{MA09\_Dnu} & $\Delta\nu$ from \citetalias{Mosser2009} pipeline & $\mathrm{\mu Hz}$ \\
		\texttt{MA09\_PARAM\_mass} & Mass from \texttt{PARAM}, based on \citetalias{Mosser2009} and APOGEE/GALAH & $\mathrm{M_{\odot}}$ \\
		\texttt{MA09\_PARAM\_rad} & Radius from \texttt{PARAM}, based on \citetalias{Mosser2009} and APOGEE/GALAH & $\mathrm{R_{\odot}}$ \\
		\texttt{MA09\_PARAM\_mbol} & Bolometric magnitude from \texttt{PARAM}, based on \citetalias{Mosser2009} and APOGEE/GALAH & $\mathrm{mag}$ \\
		\texttt{MA09\_PARAM\_dist} & Distance from \texttt{PARAM}, based on \citetalias{Mosser2009} and APOGEE/GALAH & $\mathrm{pc}$ \\
		\texttt{MA09\_PARAM\_Av} & Visual extinction from \texttt{PARAM}, based on \citetalias{Mosser2009} and APOGEE/GALAH & $\mathrm{mag}$ \\
		\texttt{MA09\_PARAM\_Ks} & $K_S$ absolute magnitude from \texttt{PARAM}, based on \citetalias{Mosser2009} and APOGEE/GALAH & $\mathrm{mag}$ \\
		\texttt{MA09\_PARAM\_parallax} & Parallax from \texttt{PARAM}, based on \citetalias{Mosser2009} and APOGEE/GALAH & $\mathrm{mas}$ \\
		\texttt{E20\_numax} & $\nu_{\mathrm{max}}$ from \citetalias{Elsworth2020} pipeline & $\mathrm{\mu Hz}$ \\
		\texttt{E20\_Dnu} & $\Delta\nu$ from \citetalias{Elsworth2020} pipeline & $\mathrm{\mu Hz}$ \\
		\texttt{E20\_PARAM\_mass} & Mass from \texttt{PARAM}, based on \citetalias{Elsworth2020} and APOGEE/GALAH & $\mathrm{M_{\odot}}$ \\
		\texttt{E20\_PARAM\_rad} & Radius from \texttt{PARAM}, based on \citetalias{Elsworth2020} and APOGEE/GALAH & $\mathrm{R_{\odot}}$ \\
		\texttt{E20\_PARAM\_mbol} & Bolometric magnitude from \texttt{PARAM}, based on \citetalias{Elsworth2020} and APOGEE/GALAH & $\mathrm{mag}$ \\
		\texttt{E20\_PARAM\_dist} & Distance from \texttt{PARAM}, based on \citetalias{Elsworth2020} and APOGEE/GALAH & $\mathrm{pc}$ \\
		\texttt{E20\_PARAM\_Av} & Visual extinction from \texttt{PARAM}, based on \citetalias{Elsworth2020} and APOGEE/GALAH & $\mathrm{mag}$ \\
		\texttt{E20\_PARAM\_Ks} & $K_S$ absolute magnitude from \texttt{PARAM}, based on \citetalias{Elsworth2020} and APOGEE/GALAH & $\mathrm{mag}$ \\
		\texttt{E20\_PARAM\_parallax} & Parallax from \texttt{PARAM}, based on \citetalias{Elsworth2020} and APOGEE/GALAH & $\mathrm{mas}$ \\
		\texttt{E17\_evstate} & \citet{Elsworth2017} evolutionary state & $\mathrm{None}$ \\
		\hline
	\end{tabular}
\end{table*}

\section{Comparison with \gaia DR3 Apsis GSP-Phot distances}
\label{app:gsp}

\begin{figure*}
	\centering
	\includegraphics[width=\hsize]{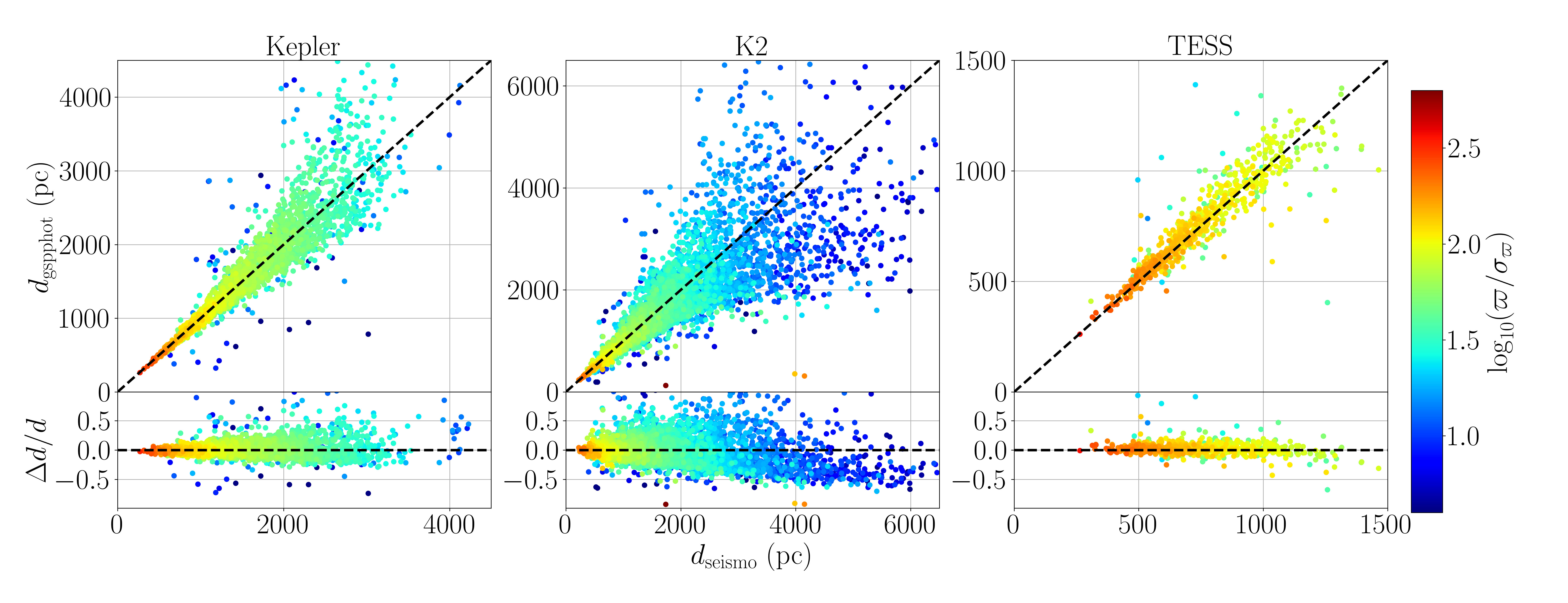}
	\caption{Comparison of \gaia DR3 Apsis GSP-Phot distances with asteroseismic ones computed using \citetalias{Elsworth2020} and APOGEE for \kepler (left), \ktwo (middle), and \tess (right). The bottom panels show the relative difference in distance. The colour scale indicates the \gaia parallax over error ratio. See Fig. 9 in \citet{Fouesneau2022} for a similar comparison with other asteroseismic datasets.}
	\label{fig:dr3_comp}
\end{figure*}

Figure \ref{fig:dr3_comp} compares distances from asteroseismology (based on \citetalias{Elsworth2020} and APOGEE) and \gaia DR3 Apsis GSP-Phot for \kepler, \ktwo, and \tess. As noted by \citet{Fouesneau2022}, a good agreement is found to about 2 kpc; beyond, GSP-Phot tends to overestimate distances as in \kepler, or on the contrary to systematically underestimate them at even further distances (see \ktwo). No issues are found for \tess nearby targets.

\section{Parallax zero-point estimates from asteroseismology}
\label{app:table}

Table \ref{table:offsets} gives a summary of the parallax offsets measured with the various combinations of asteroseismology and spectroscopy in the \kepler, \ktwo campaigns, and \tess fields.

\begin{table*}
	\caption{Parallax zero-points, measured as the difference between the \gaia EDR3 and the asteroseismic parallaxes ($\varpi_{\rm GEDR3}-\varpi_{\rm PARAM}$), for the asteroseismic fields considered in this study. Each column corresponds to a different combination of asteroseismic (\citetalias{Mosser2009} or \citetalias{Elsworth2020}) and spectroscopic (APOGEE or GALAH) constraints. The offsets reported are given in $\mu$as, and the column used in Secs. \ref{sec:results} and \ref{sec:discussion} is followed by an asterisk. The number of stars is indicated in brackets. See Sec. \ref{sec:data} for more details about the samples.}\label{table:offsets}
	\centering
	\begin{tabular}[]{c||c|c|c|c}
		Field     & \citetalias{Mosser2009}+APOGEE     & \citetalias{Elsworth2020}+APOGEE \textbf{*}    & \citetalias{Mosser2009}+GALAH & \citetalias{Elsworth2020}+GALAH   \\
		\hline
		\kepler   & $-32.02 \pm 0.36$ (4687) & $-20.09 \pm 0.33$ (4687)            & -                              & -                                 \\
		\ktwo C01 & $-18.32 \pm 3.00$ (201)             & $-21.45 \pm 3.00$ (201)   & $-14.2 \pm 2.93$ (240)         & $-14.84 \pm 2.87$ (240)           \\
		\ktwo C02 & $-22.81 \pm 2.11$ (386)             & $-22.21 \pm 2.28$ (386)   & $-28.28 \pm 1.81$ (620)        & $-28.64 \pm 1.89$ (620)           \\
		\ktwo C03 & $-8.18 \pm 2.05$ (543)              & $-6.01 \pm 2.07$ (543)    & $-3.4 \pm 2.82$ (262)          & $-5.23 \pm 2.88$ (262)             \\
		\ktwo C04 & $-14.16 \pm 1.42$ (1092)            & $-18.30 \pm 1.43$ (1092)  & $-17.68 \pm 2.48$ (474)        & $-21.39 \pm 2.55$ (474)           \\
		\ktwo C05 & $-15.50 \pm 1.41$ (865)             & $-13.59 \pm 1.46$ (865)   & $-12.46 \pm 1.65$ (757)        & $-12.39 \pm 1.65$ (757)           \\
		\ktwo C06 & $-22.04 \pm 1.56$ (690)             & $-21.20 \pm 1.60$ (690)   & $-10.82 \pm 2.72$ (262)        & $-16.51 \pm 2.53$ (262)           \\
		\ktwo C07 & $-18.29 \pm 1.91$ (422)             & $-20.03 \pm 1.99$ (422)   & $-20.67 \pm 1.42$ (822)        & $-21.41 \pm 1.46$ (822)          \\
		\ktwo C08 & $-18.78 \pm 2.24$ (436)             & $-21.11 \pm 2.35$ (436)   & $-15.7 \pm 4.0$ (159)          & $-17.65 \pm 3.89$ (159)           \\
		\ktwo C10 & $-2.08 \pm 3.59$ (199)              & $-5.65 \pm 3.49$ (199)    & $-12.58 \pm 4.07$ (117)        & $-11.07 \pm 3.99$ (117)            \\
		\ktwo C11 & $-42.82 \pm 3.56$ (189)             & $-39.15 \pm 3.66$ (189)   & $-68.18 \pm 4.79$ (235)        & $-69.92 \pm 4.81$ (235)           \\
		\ktwo C12 & $2.49 \pm 2.36$ (462)               & $1.39 \pm 2.22$ (462)     & $-10.77 \pm 6.5$ (63)          & $-10.43 \pm 6.43$ (63)             \\
		\ktwo C13 & $-30.99 \pm 2.03$ (423)             & $-33.12 \pm 2.06$ (423)   & $-37.89 \pm 2.03$ (645)        & $-37.05 \pm 1.99$ (645)           \\
		\ktwo C14 & $-28.68 \pm 2.65$ (354)             & $-26.89 \pm 2.41$ (354)   & $-23.9 \pm 4.00$ (123)         & $-16.16 \pm 4.04$ (125)           \\
		\ktwo C15 & $-11.46 \pm 13.08$ (10)             & $-7.54 \pm 11.21$ (10)    & $-14.58 \pm 1.78$ (735)        & $-7.48 \pm 1.62$ (735)            \\
		\ktwo C16 & $-23.36 \pm 2.60$ (310)             & $-21.32 \pm 2.52$ (310)   & $-13.83 \pm 3.29$ (201)        & $-8.53 \pm 3.16$ (201)           \\
		\ktwo C17 & $-17.95 \pm 2.32$ (348)             & $-18.84 \pm 2.42$ (348)   & $-11.78 \pm 3.81$ (162)        & $-11.53 \pm 3.66$ (162)           \\
		\ktwo C18 & $-10.36 \pm 5.27$ (94)              & $-2.56 \pm 5.55$ (94)     & $4.08 \pm 6.81$ (78)           & $15.23 \pm 6.35$ (78)            \\
		\tess     & $-23.23 \pm 1.27$ (1253)            & $-41.43 \pm 1.43$ (1253)  & -                              & -                     
	\end{tabular}
\end{table*}

\section{Impact of \citetalias{Lindegren2021} corrections on parallax offset estimation}
\label{app:l21_impact}

\begin{figure*}
	\centering
	\includegraphics[width=0.70\hsize]{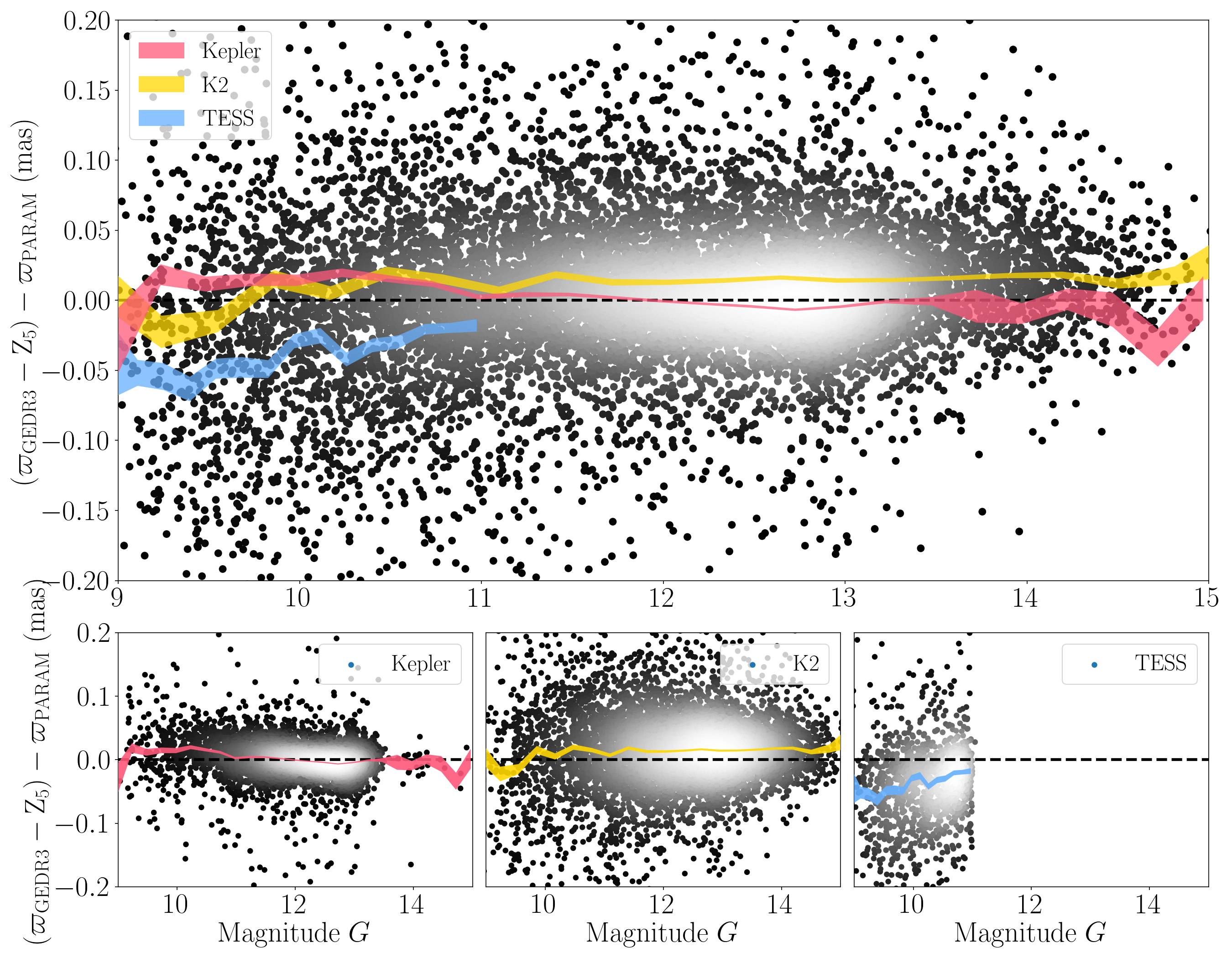}
	\caption{Same as Fig. \ref{fig:trend_G} but including \citetalias{Lindegren2021} corrections in the \gaia EDR3 parallaxes.}
	\label{fig:trend_G_L21}
\end{figure*}

\end{appendix}

\end{document}